\documentclass{article}
\usepackage{arxiv}

\usepackage[utf8]{inputenc} 
\usepackage[T1]{fontenc}    
\usepackage{hyperref}       
\usepackage{url}            
\usepackage{booktabs}
\usepackage{multirow}       
\usepackage{amsfonts}       
\usepackage{nicefrac}       
\usepackage{microtype}      
\usepackage{lipsum}		
\usepackage{graphicx}
\usepackage{natbib}
\usepackage{doi}
\usepackage{amsmath}

\newtheorem{problem}{\it{Problem}}
\newtheorem{proposition}{\it{Proposition}}
\newtheorem{proof}{\it{Proof}}

\title{Artificial Bandwidth Extension Using Deep Neural Network and $H^\infty$ sampled-data control theory}

\author{ 
	{
		Deepika Gupta}
	\\
	Department of Electronics and Electrical Engineering\\
	Indian Institute of Technology Guwahati\\
	Guwahati, Assam, India \\
	\texttt{deepika.gupta@iitg.ac.in} \\
	\And
	{Hanumant Singh Shekhawat} \\
	Department of Electronics and Electrical Engineering\\
	Indian Institute of Technology Guwahati\\
	Guwahati, Assam, India  \\
	\texttt{h.s.shekhawat@iitg.ac.in} \\
}

\renewcommand{\headeright}{}
\renewcommand{\undertitle}{}
\renewcommand{\shorttitle}{}

\begin{document}
\maketitle

\begin{abstract}
	Artificial bandwidth extension is applied to speech signals to improve their quality in narrowband telephonic communication. For accomplishing this, the missing high-frequency (high-band) components  of speech signals are recovered by utilizing a new extrapolation process based on sampled-data control theory and deep neural network (DNN).  The $H^\infty$ sampled-data control theory helps in designing of a high-band filter to recover the high-frequency signals by optimally utilizing the inter-sample signals. Non-stationary (time-varying) characteristics of speech signals forces to use numerous high-band filters. Hence, we use a deep neural network  for estimating the high-band  filter information  and a gain factor for a  specified  narrowband information of the unseen signal. The objective analysis is done on the TIMIT dataset and RSR15 dataset. Additionally, the objective analysis is performed separately for the voiced speech as well as for the unvoiced speech as generally needed in speech processing.
	Subjective analysis is done on the RSR15 dataset. 
\end{abstract}


\section{Introduction}
High  quality of speech signals in the narrowband (NB) voice communications  can be acquired by expanding the bandwidth of narrowband speech signals. It is done by  an artificial bandwidth extension (ABE) technique.  
The frequencies up to $4$ kHz  can be transmitted through the NB, i.e., $0-4$ kHz,  communication channels~\cite{li2015deep}.
Therefore, the missing high-band (HB), i.e., $4-8$ kHz,  information of the speech signals is recovered back at the recipient side by utilizing the ABE technique.
\par  Numerous ABE strategies have been proposed dependent upon the speech production model (SPM) in which the speech signal is the output of a speech production filter (SPF) (spectral envelope) driven by a residue signal~\cite{shao2005robust}. An HB residue signal and an HB spectral envelope are to be assessed by utilizing the NB signal for ABE. For accomplishing this, an HB residue signal is obtained by applying different kinds of procedures  on the narrowband residue signal,  for example,  bandpass-envelope modulated Gaussian noise~(BP-MGN)~\cite{qian2003dual}, full-wave rectification~\cite{fuemmeler2001techniques,  abel2018artificial},  pitch adaptive modulation~\cite{jax2003artificial}, spectrum folding~\cite{enbom1999bandwidth}, spectral translation~\cite{jax2003artificial,makhoul1979high}, and  harmonic noise model (HNM)~\cite{vaseghi2006speech}. An HB spectral envelope is estimated from a codebook mapping or a statistical
modeling approach by utilizing the narrowband spectral envelope.  The NB spectral envelope and the HB spectral envelope information  can be represented  by  line spectral frequencies (LSF)~\cite{li2016artificial}, linear prediction coefficients (LPC)~\cite{andersen2015bandwidth},  Mel frequency cepstral  coefficients (MFCC)~\cite{sunil2012exploration} and  linear frequency cepstral coefficients (Cepstrum)~\cite{abel2018artificial}.  
A few strategies are created  which are different from the speech production model.  In~\cite{kim2008speech}, temporal envelope and fine structure of sub-bands are taken into account  for ABE.  In~\cite{li2015deep, sadasivan2016joint, bin2015novel,bachhav2017artificial}, the WB magnitude spectrum  is taken directly for extracting beneficial HB and NB information. In~\cite{sunil2014sparse}, sparse modeling is used for  ABE, wherein it requires  a separate dictionary for voiced speech and unvoiced speech.
The HB  information for a given NB information can be assessed by the codebook mapping approaches such as vector quantization~\cite{fuemmeler2001techniques,unno2005robust}, and  the statistical modeling approaches, for example, Gaussian mixture models (GMMs)~\cite{bachhav2017artificial,pulakka2011speech,  nour2011memory,ohtani2014gmm,   li2018speech}, hidden Markov models (HMM) with GMM~\cite{abel2018artificial, seltzer2007training,  song2009study,  yaugli2013artificial}, and deep neural network (DNN) topologies~\cite{li2015deep,abel2018artificial,li2016artificial, bin2015novel,hinton2012deep, xu2014experimental, wang2015speech, abel2017dnn}.
\par As per the theory of speech production, speech production filter can be  represented accurately by a pole-zero model~\cite{marelli2010pole}. Most of the existing methods based on speech production theory,  use an  all-pole model~\cite{abel2018artificial, li2016artificial} which is  invertible. However,  it may not be sufficient  to represent envelopes of utterances like fricatives, nasals, laterals, and the burst interval of stop consonants,  since zeros present in SPF's frequency response~\cite{marelli2010pole}. In our paper,  pole-zero model is used to represent the speech production filter which may not be  invertible system~\cite{marelli2010pole}. Therefore, this problem is converted into a closed loop system which is solved using the $H^\infty$ sampled-data control theory~\cite{chen1995design, yamamoto2012signal}. It gives a robust solution in case of small modelling error~\cite{shaked1992h}.
We design a discrete filter to reconstruct the wideband speech signal based on a discrete signal model (or the speech production model in the ABE case) using the $H^\infty$ norm. The process uses available gain, phase, and inter-sample information optimally~\cite{yamamoto2012signal}.
This approach has been utilized for the audio signal in~\cite{yamamoto2012signal}, however, for a linear time-invariant (LTI) audio signal model.  It is well known that a speech signal is stationary only for short duration (approximately 10-30 ms). It means an LTI signal model is valid only for this duration. Hence,  a large number of LTI signal models and their corresponding number of reconstruction filters are essential for a better reconstruction. However, this is impractical. To this end, we build a DNN model of reconstruction filters along with NB information itself for practical usage. The GMM modeling technique is sufficient, as demonstrated in  \cite{gupta2019artificial2}. However, \cite{gupta2019artificial2} 
demands a change in the existing transmitter set-up, which is costly.
In this paper, we have proposed a new approach for synthesizing the HB signal using the $H^\infty$ sampled-data system theory and DNN (which performs better than GMM), without changing the existing transmitter set-up. Also, we use the standard technique of  gain adjustment between estimated HB signal and original HB signal, and  DFT (discrete Fourier transform) addition method for adding  NB and HB signals with slight modification~\cite{ wang2015speech, nour2008mel, abel2018simple}. 
An analysis is done by considering the two datasets using the DNN model.  

\section{A proposed set-up for ABE of speech signals}
\label{A_proposed_setup_for_ABE_of_speech_signals}
\begin{figure}[!t]
	\centering
	\includegraphics[width=4.5in]{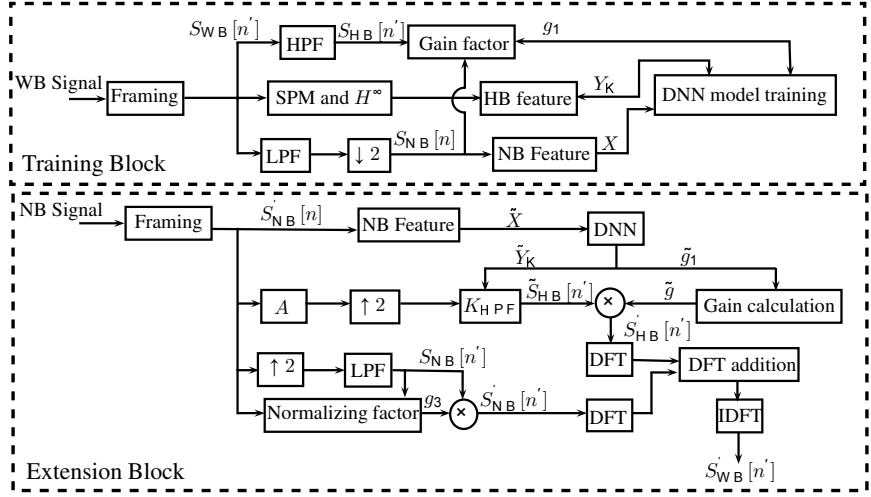}
	\caption{ Block diagram consisting the training  of the DNN model  and bandwidth extension of the narrowband  signal.}
	\label{DNN_LPC_LPF_Block_diagram_consist_training_and_extension_of_NB_signal}
\end{figure}

This section  has an explanation of  the proposed framework for the ABE. An outline of the proposed framework is shown in   Fig.~\ref{DNN_LPC_LPF_Block_diagram_consist_training_and_extension_of_NB_signal}. It  has two primary subblocks: a training block and an extension block. These subblocks are  explained within the forthcoming subsections.
\subsection{Framing} 
The fact that speech signals are non-stationary signals is well known~\cite{loizou2007speech}.  Thus, the rectangular window of 25 ms length is used to divide these signals into frames. These  frames  are supposed to be stationary signals, which allow us to find an LTI signal model for each frame. The overlap between nearby frames is settled to $50$\%.  
This process is the same in the  training block and the extension block. 
\subsection{Training Block}
The training block in   Fig.~\ref{DNN_LPC_LPF_Block_diagram_consist_training_and_extension_of_NB_signal} has two sequential processes,  one of them is to extract features for training, and another is to train the DNN model~\cite{abel2018artificial} by utilizing the training features.  
\subsubsection{Feature extraction}
\begin{figure}[t]
	\centerline{
		\includegraphics[width=3.0in,height=.45in] {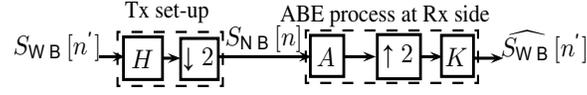}}
	\caption{A block for the producing the NB signal and reconstruction of a wideband signal.}
	\label{LPF_LPC_letter_Set-up for Reconstruction}
\end{figure}

Feature extraction  is performed frame-wise. First, an HB feature is computed using the WB stationary signal ($S_{WB}[n']$). For this, we join the transmitter (Tx) set-up and ABE process used at the receiver (Rx) side in Fig.~\ref{LPF_LPC_letter_Set-up for Reconstruction} for a single frame.   
In Fig.~\ref{LPF_LPC_letter_Set-up for Reconstruction},  $\downarrow$ $2$ depicts  a downsampler with downsampling factor $2$, $H$ is the non-causal FIR (finite impulse response) low pass filter, $\uparrow$ $2$ depicts an upsampler with upsampling factor $2$, and filter $A$ is the reciprocal of the  all-pole model of  signal $S_{NB}[n]$ obtained by linear prediction (LP) analysis~\cite{makhoul1975linear}. An output of the filter $A$ is the narrowband residue signal.   $S_{NB}[n]$ is the narrowband signal, and $\widehat{S_{WB}}[n']$ is the estimated wideband signal, with $n$ and $n'$ are the 8 kHz and 16 kHz sample index, respectively.  Further, we proposed an error set-up in Fig.~\ref{LPF_LPC_letter_Error_System} for finding the filter $K$, wherein we extract the wideband information in terms of the speech production model (signal model) $F$. We obtained  the signal model $F$ using Prony's method~\cite{markel2013linear}. { The number of poles and zeros was empirically calculated for
	each frame in such a way that the error is minimized.  $S_{WB}[n']$ is the output of signal model $F$ driven by  the white noise for unvoiced speech or an impulse train for voiced speech.    
	\begin{figure}[!t]
		\centerline{
			\includegraphics[width=3.89in, height=.64in]
			{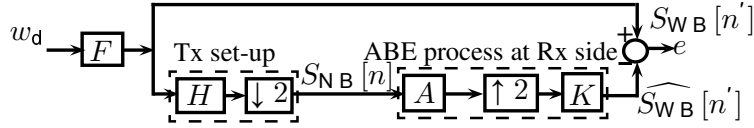}}
		\caption{Proposed  an error set-up  for a WB  signal reconstruction.}
		\label{LPF_LPC_letter_Error_System}
	\end{figure} 
	Here, a filter $K$ is obtained by minimizing the error $e$ by the following optimization problem. 
	\begin{problem}%
		\label{problem_LPC}
		Given a speech production model (signal model) $F$ and a   low pass filter $H$,  design a stable and causal  synthesis filter $K_{opt}$ defined as
		\begin{align*}
		K_{opt}:=&\arg \min_{K}( ||\mathbb{W}||_\infty ),
		\end{align*}
		where $\mathbb{W}$ is the discrete error system defined as
		\begin{align}
		\mathbb{W}:=  F- K(\uparrow 2)A(\downarrow 2) H F,
		\label{Problem_form_eqn_G_1}
		\end{align} 
		with input $w_d$ (either white noise for the unvoiced speech or quasi-periodic impulse train for the voiced
		speech) and output $e$ (see Fig.~\ref{LPF_LPC_letter_Error_System}).  $||\mathbb{W}||_\infty$ denotes the $H^\infty$-norm of the system $\mathbb{W}$, which is  defined as ( see~\cite{ yamamoto2012signal}),
		\begin{align*}
		||\mathbb{W}||_\infty:=\sup_{w_d\neq 0} \frac{||e||_2 }{||w_d||_2}.
		\end{align*}	
	\end{problem} 
	
	Further,  problem~\ref{problem_LPC} can be reformulated  as an $H^\infty$ sampled-data control problem~\cite{chen2012optimal}. The $H^\infty$ norm is generally preferred  in case of the signal modeling error~\cite{shaked1992h}.
	The solution is essentially given in Appendix A (see also~\cite{chen1995design, yamamoto2012signal, gupta2019artificial2,gupta2019artificial}). The obtained infinite impulse response  filter $K$
	consists the NB information and HB information as well. Here, we need to estimate the HB information of an WB signal only. So, filter $K$ is cascaded with a causal FIR high pass filter for extracting  its HB information only. As a result, we obtain a new  HB IIR filter $K_{\text{HPF}}$.   The obtained HB IIR filter is converted into a finite impulse response filter by truncating its Taylor series. We did the FIR approximation of IIR filter $K_{\text{HPF}}$ empirically using a process similar  to~\cite{gupta2019artificial2,gupta2019artificial}. Later on, an order of  FIR filter is fixed to 21.  In essence, this FIR filter consists the high-band spectral envelope information. This FIR approximation of IIR filter $K_{\text{HPF}}$  is  used as the HB feature vector. 
	Narrowband feature vector is representing  the linear prediction coefficients  (LPC)~\cite{aida2006investigation},  which are calculated  by using the  signal ($S_{NB}[n]$).  The dimension of the NB feature vector  is settled to  eleven. }
\par For gain adjustment, energy of the estimated HB signal is set  equal to  the energy of  original HB signal. For this,  the gain factor is calculated as
\begin{align}
g_1=\log 10 \frac{\sum_{n'=1}^{N}S_{HB}[n']^2}{\sum_{n=1}^{N/2}S_{NB}[n]^2},
\end{align}
where $S_{HB}[n']$ signifies  the HB  signal acquired by high pass filtering of the signal $S_{WB}[n']$, and $N$ is the signal $S_{HB}[n']$ length.
\subsubsection{ DNN model}
\label{ DNN_model}
After extracting  the  features for training, 
the DNN model is trained  using the NB feature vector  as an input vector $X\in \mathbb{R}^{11}$, and the concatenated HB feature vector $Y_K\in \mathbb{R}^{21}$ and gain factor $g_1$ as an output vector  with the mean square error being considered as a loss function (see DNN-R in~\cite{abel2018artificial}). The mean and variance normalization (MVN) has also applied to both the input vector and the output vector~\cite{li2015deep}. 
\subsection{Extension Block}
After training the DNN model, WB signal is assessed in the extension block, which has three main parallel procedures, as  explained briefly in the next sections.
\subsubsection{ High-band feature and gain estimation}
Firstly, the NB feature vector is computed for a given  NB signal $S^{'}_{NB}[n]$.  MVN is  applied to the NB feature vector and then fed to the DNN. A reverse MVN procedure  is applied to the  DNN output~\cite{li2015deep}.  As a result, we acquire the estimated HB feature vector ($\tilde{Y}_K$) and gain $\tilde{g}_1$. 
\subsubsection{ High-band signal estimation} 
In the initial step, we calculate the filter $A$ for a given NB signal $S^{'}_{NB}[n]$. 
The NB signal $S^{'}_{NB}[n]$ is now passed through the filter $A$ and then upsampled by a factor of $2$. The resulting signal is then fed via the  estimated HB filter $K_{\text{HPF}}$, which gives an intermediate HB signal $\tilde{S}_{HB}[n']$. Next, the synthesized energy  of  $\tilde{S}_{HB}[n']$ is made approximately  equivalent to the energy of the corresponding original HB signal by introducing gain adjustment as
\begin{align}
S^{'}_{HB} [n']=\tilde{g}~\tilde{S}_{HB}[n'], 
\end{align} 
where
\begin{align*}
\tilde{g}=\sqrt{\frac{10^{\tilde{g}_1}}{10^{\tilde{g}_2}}}, \text{with}~~ \tilde{g}_2=\log 10  \frac{\sum_{n'=1}^{N}\tilde{S}^2_{HB}[n']}{\sum_{n=1}^{N/2}{S^{'}}^2_{NB}[n]}.
\end{align*} 
$S^{'}_{HB} [n']$ is the estimated HB signal. 
\subsubsection{ Narrowband signal processing}
The NB signal $S_{NB}[n']$  is accessed by passing the NB signal $S^{'}_{NB}[n]$ through the $\uparrow 2$ following the filter $H$. Then, the following equation calculates a normalization factor $g_3$ as~\cite{gupta2019artificial}
\begin{align*}
g_3= \frac{\max(S^{'}_{NB}[n])}{\max(S_{NB}[n'])}.
\end{align*} 
The NB signal $S^{'}_{NB}[n']$ is then calculated as
\begin{align*}
S^{'}_{NB}[n']=g_3 S_{NB}[n'].
\end{align*}
\subsubsection{ Wideband signal estimation}
{ The estimated HB signal $S^{'}_{HB} [n']$  and  the NB signal $S^{'}_{NB}[n']$ are not added directly because the estimated HB signal $S^{'}_{HB} [n']$ contains some information in its narrowband region. The  filter $K_{\text{HPF}}$ is not  ideal HB filter. Somehow, it passes  some amount of NB information.    Thereby,  we add them in the frequency domain using the DFT~\cite{abel2018simple}. For this, we take $N$ point DFT of both signals. The DFT addition is applied to them and  computed as 
	\begin{align*}
	S^{'}_{WB}[k]=\begin{Bmatrix}
	S^{'}_{NB}[k],& k=0,1,..,\frac{N}{4}\\ 
	S^{'}_{HB} [k], & k=\frac{N}{4}+1,..,\frac{N}{2}+\frac{N}{4}-1\\
	S^{'}_{NB}[k], & k=\frac{N}{2}+\frac{N}{4}.,N-1
	\end{Bmatrix}
	\end{align*}
	where the resulting $S^{'}_{WB}[k]$ is the DFT of an estimated wideband signal. 
	Here, $S^{'}_{NB}[k]$ and $S^{'}_{HB} [k]$  are the DFTs of  $S^{'}_{NB}[n']$ and $S^{'}_{HB} [n']$ signals, respectively.  Lastly, the inverse DFT (IDFT) of $S^{'}_{WB}[k]$ is computed giving the estimated wideband signal ($S^{'}_{WB}[n']$).  
	An advantage of this process is no over-estimation in any of the bands. Later, it can be  seen its effect on performances in Section~\ref{Results_LPF_LPC_HPF}.}
\section{Experimental Set-up and Results}
\label{Experimental_Analysis_and_Results}
\subsection{Experimental Set-up}
We analyze the performances  for the TIMIT dataset \cite{garofolo1993timit} and the RSR15 dataset~\cite{larcher2014text}.   Both the datasets contain the recorded speech files at a sampling rate of $16$ kHz.   TIMIT dataset is already segmented into two distinct sets as a testing set and a training set. We truncate these sets so that  new truncated sets have the same percentage of female and male speech files. Then, the truncated train and  test sets are taken as a training set and a validation set, respectively. Then, performances are evaluated on the speech files belonging to  the RSR15 dataset. Both  causal high pass filter (HPF) and non-causal low pass filter (LPF) are the same  as characterized  in~\cite{gupta2019artificial}.  
\subsection {Results}
\label{Results_LPF_LPC_HPF}
Objective and subjective assessments are carried out  to  analyze the quality  of the artificially extended speech signals. For this purpose, segSNR (segmental signal to noise ratio)~\cite{hu2008evaluation}, objective metrics  MOS-LQO (mean opinion score listening quality objective)~\cite{itu2003862},  and LSD (log spectral distance for  WB)~\cite{abel2018simple}, and  subjective measure MOS-LQS (subjective listening quality mean opinion score)~\cite{malfait2006p} are chosen. Further, experiments are carried out on the validation set for the different actions such as the NB processing, gain adjustment, and the  DFT addition in the condition of utilizing the filter $K_{\text{HPF}}$ directly (oracle $K_{\text{HPF}}$) in Fig.~\ref{DNN_LPC_LPF_Block_diagram_consist_training_and_extension_of_NB_signal}. It is done to investigate  the performances acquired by the proposed filter $K_{\text{HPF}}$ under two conditions: IIR and FIR with 21 coefficients. 
As we see the performances for each significant output such as $S_{NB}[n']$, $S^{'}_{NB}[n']$, time added output without gain adjustment,  time added output with gain adjustment, i.e.,  $S^{'}_{NB}[n']+S^{'}_{HB}[n']$, and DFT addition output with gain adjustment, i.e.,  $S^{'}_{WB}[n']$, on the validation set  in Table~\ref{dirct_k_hpf_used_check_perform_on_testset}.
\begin{table}[t]
	\centering
	\caption{Performances computed on the validation set  in case of  use the filter $K_{\text{HPF}}$ directly, i.e., oracle $K_{\text{HPF}}$, in Fig.~\ref{DNN_LPC_LPF_Block_diagram_consist_training_and_extension_of_NB_signal} for ABE.}
	\resizebox{.83\textwidth}{!}{
		\begin{tabular}{|c|c|c|c|c|c|c|c|c|}
			\hline
			\multirow{2}{*}{Conditions} & \multirow{2}{*}{$S_{NB}[n']$} & \multirow{2}{*}{$S^{'}_{NB}[n']$} & \multicolumn{2}{c|}{\begin{tabular}[c]{@{}c@{}}Time addition\\ without gain\end{tabular}} & \multicolumn{2}{c|}{\begin{tabular}[c]{@{}c@{}}Time addition\\ with Gain\end{tabular}} & \multicolumn{2}{c|}{\begin{tabular}[c]{@{}c@{}}DFT addtion\\ with Gain\end{tabular}} \\ \cline{4-9} 
			&    & & $K_{\text{HPF}}$ (IIR)  & $K_{\text{HPF}}$ (FIR)  & $K_{\text{HPF}}$ (IIR) & $K_{\text{HPF}}$ (FIR)  & $K_{\text{HPF}}$ (IIR) & $K_{\text{HPF}}$ (FIR)  \\ \hline
			LSD  & 11.2339 & 9.5331& 8.7417  & 8.7201& 7.3694& 7.4207  &   6.8355 &6.7871    \\ \hline
			segSNR  & 4.83& 14.70  & 13.18  & 13.36  & 13.18  & 13.06  & 13.76   & 13.70     \\ \hline
			MOS-LQO   & 3.8468   & 3.6932  & 3.6889    & 3.6832     & 3.6833    & 3.6062  &  3.9137  &   3.9185   \\ \hline
	\end{tabular}}
	\label{dirct_k_hpf_used_check_perform_on_testset}
\end{table}
As evident from  Table~\ref{dirct_k_hpf_used_check_perform_on_testset}, the gain adjustment improves  the LSD significantly applied to the time addition case. Then, the DFT addition improves  the all objective measures more than the time addition by considering the gain adjustment for both additions.        Next, we  proceed to measure the performances for  the proposed approach using the DNN  model.      
\subsubsection{Objective assessment} 
Here, we do tests to decide an optimal DNN architecture for the proposed HB feature and the LPC NB feature.  For deciding the DNN architecture, hyper-parameters  such as the batch size, epoch, and the learning rate  are fixed to $300$, $50$, and $0.01$, respectively.  
Different DNN architectures are trained using the training set features over the fixed AdaMax (adaptive moment estimation based on the infinity norm)~\cite{kingma2014adam} optimizer,  ReLU activation function in hidden layers, linear activation function in the output layer, and LPC NB feature definition.  Batch normalization before activation function and $L^2$-regularization are utilized as well.  The decay rates $\beta_1$ for the first-moment estimate and $\beta_2$ for the second-moment estimate are set at $0.9$ and $0.999$ for the AdaMax optimizer, respectively.   Then, different DNN topologies obtained by changing the number of hidden layers ($N_{HL}$) and the number of neurons ($N_U$) in each hidden layer are trained.  These topologies  are tested   by computing the LSD on the validation set, as tabulated in Table~2.
\begin{table}[]
	\centering
	\caption{ LSD computed on the validation set by changing $N_{HL}$  and $N_U$ for the fixed batch size $300$ and ReLU activation function in hidden layers.}
	\resizebox{.47\textwidth}{!}{	
		\begin{tabular}{|c|c|c|c|c|c|}
			\hline
			$N_{HL}$ & $N_U$ & LSD & $N_{HL}$ & $N_U$ & LSD \\ \hline
			3 & 128 &7.1899  & 	4 & 1024 & 7.1478 \\ \hline
			3 & 256 &7.1866 &5 & 128 &  7.1842\\ \hline
			3 & 512  & 7.1801  & 5& 256&  7.1518\\ \hline
			3 & 1024 & 7.1723 &5& 512&7.1638  \\ \hline
			4 & 128 &7.1582 &  5& 1024& 7.1485 \\ \hline
			4 & 256 &   7.1511 & 6 & 128&  7.1500  \\ \hline
			\bf 4 & \bf 512 &  \bf 7.1448 &6 & 256 & 7.1661 \\ \hline
	\end{tabular}}
	\label{varying_DNN_performance_Evaluation_LPC_LPF_hpf}
\end{table}       
In Table~\ref{varying_DNN_performance_Evaluation_LPC_LPF_hpf}, we see that increasing the number of hidden layers ($N_{HL}$) are influencing LSD more than  increasing the number of neurons ($N_U$) in each hidden layer. We pick a topology for DNN with $4$ $N_{HL}$ and $512$  $N_U$.  Furthermore, this architecture is tuned  for  different sizes, and then batch size $180$, producing minimum LSD, is selected.  Thus, we use an optimal DNN topology planned with the ReLU activation function in hidden layers, $180$ batch size, four hidden layers, $512$ units in each hidden layer for inspecting the test set performance.  Then, objective  metrics are arranged in Table~3 for the artificially extended speech files belonging to the test set by the proposed framework using DNN model  also, GMM model~\cite{kain1998spectral}, the modulation technique~\cite{abel2018artificial} with utilizing gain defined in~\cite{nour2008mel}, cepstral domain approach~\cite{abel2018simple} and  the spectrum folding technique utilizing gain defined in~\cite{wang2015speech}  with keeping the same experimental conditions such as window duration, type of window, NB processing, dimensions of HB feature and NB feature, and DNN model. But, the input feature dimension is required high for the cepstral domain approach, as see~\cite{abel2018simple}. The LSF feature is used to represent the NB and HB features for the modulation technique and spectrum folding technique.     
\begin{table}[t]
	\centering
	\caption{Performances figured out on the testing set (RSR15) for the proposed approach and the existing approaches.}
	\resizebox{0.7\textwidth}{!}{
		\begin{tabular}{|c|c|c|c|c|c|c|}
			\hline
			Method  & MOS-LQO  & LSD & segSNR  \\
			\hline
			Proposed approach with $128$ GMMs &  3.6186  &  8.3984   &12.3040  \\ \hline
			Proposed approach with DNN model &3.7474   & 8.1927 &  12.3463 \\
			\hline
			Spectrum folding approach with DNN model &3.4693  &  8.5207 & 11.9991\\
			\hline
			Modulation technique with DNN model &3.2478  & 8.7193  &11.1167\\ 	\hline
			Cepstral Domain with DNN model & 3.7018  &  8.5566 &13.3632\\
			\hline
	\end{tabular}}
	\label{fixed_DNN_performance_Evaluation_LPC_LPF_spetrum_folding}
\end{table}
As viewed in Table~\ref{fixed_DNN_performance_Evaluation_LPC_LPF_spetrum_folding}, the proposed approach  improves  MOS-LQO and LSD significantly, whereas the cepstral domain approach gives better  segSNR in comparison to the other approaches.
\par Additionally, we figure out the performances for two  kinds of speech signals: voiced speech and  unvoiced speech.  These signals are extracted from the test set speech signals using the glottal activity detection  (GAD) technique~\cite{adiga2015detection}. Performances are listed in Table~\ref{fixed_DNN_performance_Evaluation_LPC_LPF_voiced}  for the voiced speech and Table~\ref{fixed_DNN_performance_Evaluation_LPC_LPF_unvoiced} for  the unvoiced speech  obtained by the proposed approach and the existing approaches.      
\begin{table}[t]
	\centering
	\caption{Performances figured out for  voiced speech extracted from the speech files belonging to  the test set  }
	\resizebox{0.7\textwidth}{!}{
		\begin{tabular}{|c|c|c|c|c|c|c|}
			\hline
			Conditions  & MOS-LQO  & LSD & segSNR  \\ 
			\hline
			Proposed approach  using	128 GMMs &  4.2663  & 7.9297   &  16.3455 \\ \hline
			Proposed approach  using	DNN model  &  4.2666  &    7.6227 &   16.7151
			\\
			\hline
			Spectrum folding approach with DNN model &  4.1630 &8.0643   &15.5798 \\
			\hline
			Modulation technique with DNN model &3.9243 &  7.8000 &15.2600\\ 	\hline
			Cepstral Domain with DNN model & 4.2496  & 7.5966 &17.7857\\
			\hline						
	\end{tabular}}
	\label{fixed_DNN_performance_Evaluation_LPC_LPF_voiced}
\end{table}      
\begin{table}[t]
	\centering
	\caption{ Performances figured out  for unvoiced speech extracted from the speech files belonging to  the test set}
	\resizebox{0.7\textwidth}{!}{
		\begin{tabular}{|c|c|c|c|c|c|c|}
			\hline
			Conditions  & MOS-LQO  & LSD & segSNR  \\ \hline
			Proposed approach using	128 GMMs & 4.2485
			&  8.3967 & 10.3930 \\ \hline
			
			Proposed approach using DNN model  & 4.2514    &  8.2123    & 10.4058  \\
			\hline
			Spectrum folding approach with DNN model &4.1977  &  8.5109 & 10.3963 \\
			\hline
			Modulation technique with DNN model &3.9289 &8.6697   &9.5482\\ 	\hline
			Cepstral Domain with DNN model & 4.2143  &9.1502  &11.3974\\
			\hline
	\end{tabular}}
	\label{fixed_DNN_performance_Evaluation_LPC_LPF_unvoiced}	
\end{table}      
All the objective measures are improved  by the proposed approach using the DNN model for both kinds of  speech in comparison to the  GMM model. LSD is significantly decreased by 0.30 dB for voiced speech and 0.18 dB for the unvoiced speech by the proposed approach using the  DNN model with respect to the GMM model.  segSNR is high for the  voiced speech because of their  high  energy in comparison to the unvoiced speech. It is a verifiable truth that voiced speech and unvoiced speech energies lie a lot of in NB and HB, respectively. Therefore, measured LSD is lower for the voiced speech than the unvoiced speech. segSNR is obtained comparatively better for the cepstral domain approach than the other approaches for both types of  speech. Rest of the measures are obtained better by the proposed approach using the DNN model.  
\subsubsection{Subjective assessment}
In a typical telephonic conversation, the perceptual quality of the receiving speech signal has given more priority. For this,  
subjective assessment is done by following~\cite[Annex E] {rec1996p}. A CMOS score is computed by averaging the  score between $-3$ (much worse) to $3$ (much better) given by twelve speakers for the randomly selected twenty artificially extended speech files. Speakers are naive for this assessment.  This score is calculated for the three conditions in which the artificially extended speech files  by the proposed approach are compared to the  artificially extended speech files by the three existing approaches using the same DNN model. 
\begin{table}[!t]
	\centering
	\caption{{Subjective assessment evaluated  on artificially extended speech files  belonging to the test set using the  DNN model }}
	\label{CMOS_spectrum_fold_proposed_GMM_DNN}
	\resizebox{0.53\textwidth}{!}{
		\begin{tabular}{|c|c|}
			\hline
			Conditions & CMOS \\ \hline
			Spectrum folding vs Proposed approach  & 0.6150  \\ \hline
			Modulation technique vs Proposed approach  & 0.6450 \\ \hline
			Cepstral Domain  vs Proposed approach & 0.5950 \\ \hline	
	\end{tabular}}
\end{table}
As evident in Table~\ref{CMOS_spectrum_fold_proposed_GMM_DNN}, CMOS is improved by $0.6150$, $0.6450$, and $0.5950$ value for the proposed approach in comparison to the spectrum folding approach, modulation technique, and cepstral domain approach using the same DNN model, respectively. 
\section{Conclusion}          
This work proposes a new ABE approach for speech signals that uses the  (discrete domain) $H^\infty$ sampled-data control theory and a DNN model along with well-known techniques of gain adjustment and the DFT addition with a slight modification. The $H^\infty$ optimization and the speech production model help in acquiring a high-band filter, which takes the inter-sample signal, the signal model's gain and phase characteristics into account. Objective and subjective measures used in our analysis are improved by the proposed approach in comparison to the existing approaches using the same DNN  model. This paper shows the potential of the sampled-data system in speech processing with a lot of possibilities for further research.

\section*{Appendix A}
\label{appendix}
\begin{figure}[!h]
	\centerline{
		\includegraphics[width=3in,height=1in]{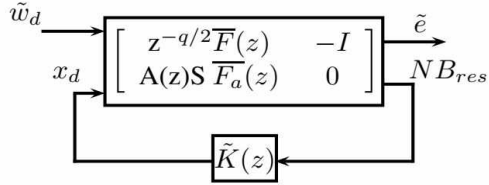}}
	\caption{ Standard feedback control system}
	\label{LPF_LPC_Standard_Control_system_2}
\end{figure}

The error system in Fig.~\ref{LPF_LPC_letter_Error_System}  is a multi-rate system that can be converted into a single rate system using the lifting technique~\cite{chen1995design,chen2012optimal}. The z-transform representations of lifting and inverse lifting~\cite{yamamoto2012signal} are
\begin{subequations}
	\begin{align}
	{\bf L_N} &= (\downarrow N)\begin{bmatrix}1&z&z^2&.....&z^{N-1}\end{bmatrix}^T, \label{lifting} 
	\\{\bf L_N^{-1}}&=\begin{bmatrix}1&z^{-1}&z^{-2}&.....&z^{-(N-1)}\end{bmatrix}(\uparrow N). \label{inverse_lifting}
	\end{align}
\end{subequations} 
\textit{
	\begin{proposition}
		Let  the transfer function  $G(z)$ be represented in state space as
		\begin{align*}
		G(z):=\left[ \begin{array}{c|c} {\bf A}&{\bf B}\\\hline {\bf C}&{\bf D} \end{array}\right]={\bf D}+{\bf C}(z{\bf I-A})^{-1}{\bf B},
		\end{align*}
		with ${\bf A}\in \mathbb{R}^{N\times N}, {\bf B}\in \mathbb{R}^{N\times p}, {\bf C}\in \mathbb{R}^{m\times N}, {\bf D}\in \mathbb{R}^{m\times p}$ matrices, $m$ and $p$ being the dimensions of output and input of $G(z)$, respectively. Next,  the lifted (by a factor of 2) transfer function of $G(z)$  in state space form is represented as
		\begin{align}
		\overline{G}(z):={\bf L_2}G(z){\bf L_2^{-1}}=\left[ \begin{array}{c|c} {\bf A}^2&\begin{matrix}{\bf A} {\bf B}&{\bf B}&\end{matrix}
		\\\hline\begin{matrix}{\bf C}\\{\bf C} {\bf A}\end{matrix}&\begin{matrix}{\bf D}&{\bf 0}\\{\bf CB} & {\bf D} \end{matrix} \end{array}\right],
		\label{lifted_transfer_G}
		\end{align}
		where $\bf L_2$ and ${\bf L_2^{-1}}$ can be obtained by using \eqref{lifting} and \eqref{inverse_lifting} respectively.
\end{proposition}} 
\begin{proof}
	See ~\cite[Theorem 8.2.1]{chen2012optimal}  .
\end{proof}
We use \eqref{lifted_transfer_G} to  get the following results from~\cite{yamamoto2012signal}
\begin{align}
K(z)(\uparrow 2)&={\bf L_2^{-1}}\tilde{K}(z),\nonumber  \\
K(z)&=\begin{bmatrix}1&z^{-1}\end{bmatrix}\tilde{K}(z^2), \label{processing_filter}
\end{align}
where $\tilde{K}(z):=\overline{K}(z)\begin{bmatrix} 1&0\end{bmatrix}^T_{1\times 2}$ and $\overline{K}(z):= {\bf L_2}K(z) {\bf L_2^{-1}}$. The error system $\mathbb{W}$  defined in \eqref{Problem_form_eqn_G_1} can be written in z-domain 
\begin{align}
\mathbb{W}(z):=  F(z)- {\bf L_2^{-1}}\tilde{K}(z)A(z)(\downarrow 2) H(z) F(z).
\label{lpf_lpc_G_equn1}
\end{align} 
Here, $H(z)$ is a non-causal and zero phase  FIR low pass filter.
Let  $H(z)$ be represented as
\begin{align}
H(z)=&a_{q}z^{-q}+..+a_{1}z^{-1}+a_0+ a_1 z^1+...+a_qz^q, \nonumber
\\ =&z^q H_{\text{causal}}(z),
\label{H_Z_causal_eqn} 
\end{align}
where $H_{\text{causal}}(z):=(a_{q}z^{-2q}+..+a_{1}z^{-(q+1)}+a_0z^{-q}+ a_1 z^{-(q-1)}+...+a_q)$ with $a_i\in \mathbb{R}$ and $q$ can be assumed  even integer number without loss of generality. We substitute  $H(z)$ from~\eqref{H_Z_causal_eqn} into~\eqref{lpf_lpc_G_equn1} as 
\begin{align}
\mathbb{W}(z)=& F(z)- {\bf L_2^{-1}} \tilde{K}(z) A(z) (\downarrow 2) z^q H_{\text{causal}}(z) F(z), \nonumber
\\=&F(z)- {\bf L_2^{-1}} \tilde{K}(z) A(z)z^{q/2} (\downarrow 2)  
F_a(z),
\end{align} 
where $F_a(z):=H_{\text{causal}}(z) F(z)$. $F_a(z)$ and $\tilde{K}(z)A(z)z^{q/2}$ are the transfer functions at different sampling rate. Hence, we lift the input and output of $\mathbb{W}$   by a factor of 2. It gives the lifted transfer function of  $\mathbb{W}$  as follows
\begin{align*} 
\overline{\mathbb{W}}(z)&= {\bf L_2} \mathbb{W}(z) {\bf L_2^{-1}},\nonumber
\\&= \overline{F}(z) - \tilde{K}(z) A(z)z^{q/2} S \overline{F_a}(z), 
\end{align*}
with  $ \overline{F}(z):= {\bf L_2} F(z) {\bf L_2^{-1}}$,  $S=\begin{bmatrix} 1&0 \end{bmatrix}$, and $\overline{F_a}(z):= {\bf L_2} F_a(z) {\bf L_2^{-1}}$. After applying lifting to the multi rate system $\mathbb{W}$, we obtain a single rate system $\overline{\mathbb{W}}$ which  is a non-causal system due to the presence of $z^{q/2}$. $\overline{\mathbb{W}}$ needs to convert into a causal system to utilize the robust control tool in MATLAB~\cite{chiang1997matlab}. Therefore, we make it causal by multiplication  with $z^{-q/2}$ in the subsequent step, 
\begin{align}
\mathcal{W}=& z^{-q/2} \overline{\mathbb{W}}, \nonumber
\\ =&z^{-q/2} \overline{F}(z) - \tilde{K}(z) A(z) S \overline{F_a}(z).
\label{lpf_lpc_last_eqn_system}
\end{align}
Finally, we get a causal as well as single rate system $\mathcal{W}$. Now, original system $\mathbb{W}$ has been converted  into $\mathcal{W}$. However, $H^\infty$-norm of these modified systems remain the same as the original system, i.e., $||\mathcal{W}||_\infty=||\overline{\mathbb{W}}||_\infty=||\mathbb{W}||_\infty$~\cite{chen2012optimal}. The $H^\infty$-norm of the  system $\mathcal{W}$ is minimized  by using an optimal filter $\tilde{K}(z)$. \eqref{lpf_lpc_last_eqn_system} is written in the form of a standard feedback control system (closed loop system) as depicted in Fig.~\ref{LPF_LPC_Standard_Control_system_2}~\cite{chen2012optimal}.
Here, $NB_{res}$ denotes the NB residual, $\bf 0$ is a zero matrix of  $1 \times 2$, $I$ is an identity matrix of $2 \times 2$, $\tilde{w}_d = {\bf L_2} w_d $, and $\tilde{e} ={\bf L_2}e$. Further, we obtain an optimal filter $\tilde{K}(z)$ with the help of robust control toolbox in MATLAB~\cite{chiang1997matlab}. To this end, an  optimal synthesis filter $K(z)$ is obtained from  $\tilde{K}(z)$ by using~\eqref{processing_filter}.

\bibliographystyle{unsrtnat}
\bibliography{references}  

\begin{thebibliography}{52}
\providecommand{\natexlab}[1]{#1}
\providecommand{\url}[1]{\texttt{#1}}
\expandafter\ifx\csname urlstyle\endcsname\relax
  \providecommand{\doi}[1]{doi: #1}\else
  \providecommand{\doi}{doi: \begingroup \urlstyle{rm}\Url}\fi

\bibitem[Li and Lee(2015)]{li2015deep}
Kehuang Li and Chin-Hui Lee.
\newblock A deep neural network approach to speech bandwidth expansion.
\newblock In \emph{Proceedings IEEE International Conference on Acoustics,
  Speech and Signal Processing (ICASSP)}, pages 4395--4399. IEEE, 2015.

\bibitem[Shao(2005)]{shao2005robust}
Xu~Shao.
\newblock \emph{{Robust Algorithms for Speech Reconstruction on Mobile
  Devices}}.
\newblock PhD thesis, University of East Anglia, 2005.

\bibitem[Qian and Kabal(2003)]{qian2003dual}
Yasheng Qian and Peter Kabal.
\newblock Dual-mode wideband speech recovery from narrowband speech.
\newblock In \emph{Eighth European Conference on Speech Communication and
  Technology, GENEVA, Switzerland}, pages 1433--1436, 2003.

\bibitem[Fuemmeler et~al.(2001)Fuemmeler, Hardie, and
  Gardner]{fuemmeler2001techniques}
Jason~A Fuemmeler, Russell~C Hardie, and William~R Gardner.
\newblock Techniques for the regeneration of wideband speech from narrowband
  speech.
\newblock \emph{EURASIP Journal on Applied Signal Processing}, 2001\penalty0
  (1):\penalty0 266--274, 2001.

\bibitem[Abel and Fingscheidt(2018)]{abel2018artificial}
Johannes Abel and Tim Fingscheidt.
\newblock {Artificial Speech Bandwidth Extension Using Deep Neural Networks for
  Wideband Spectral Envelope Estimation}.
\newblock \emph{IEEE/ACM Transactions on Audio, Speech, and Language
  Processing}, 26\penalty0 (1):\penalty0 71--83, 2018.

\bibitem[Jax and Vary(2003)]{jax2003artificial}
Peter Jax and Peter Vary.
\newblock On artificial bandwidth extension of telephone speech.
\newblock \emph{Signal Processing}, 83\penalty0 (8):\penalty0 1707--1719, 2003.

\bibitem[Enbom and Kleijn(1999)]{enbom1999bandwidth}
Niklas Enbom and W~Bastiaan Kleijn.
\newblock Bandwidth expansion of speech based on vector quantization of the mel
  frequency cepstral coefficients.
\newblock In \emph{Proceedings IEEE Workshop on Speech Coding}, pages 171--173.
  IEEE, 1999.

\bibitem[Makhoul and Berouti(1979)]{makhoul1979high}
John Makhoul and Michael Berouti.
\newblock High-frequency regeneration in speech coding systems.
\newblock In \emph{Proceedings IEEE International Conference on Acoustics,
  Speech, and Signal Processing, Cambridge, United Kingdom}, volume~4, pages
  428--431. IEEE, 1979.

\bibitem[Vaseghi et~al.(2006)Vaseghi, Zavarehei, and Yan]{vaseghi2006speech}
Saeed Vaseghi, Esfandiar Zavarehei, and Qin Yan.
\newblock Speech bandwidth extension: {Extrapolations} of spectral envelop and
  harmonicity quality of excitation.
\newblock In \emph{Proceedings International Conference on Acoustics, Speech
  and Signal Processing (ICASSP)}, volume~3, pages III--844--III--847. IEEE,
  2006.

\bibitem[Li and Kang(2016)]{li2016artificial}
Yaxing Li and Sangwon Kang.
\newblock Artificial bandwidth extension using deep neural network-based
  spectral envelope estimation and enhanced excitation estimation.
\newblock \emph{IET Signal Processing}, 10\penalty0 (4):\penalty0 422--427,
  2016.

\bibitem[Andersen et~al.(2015)Andersen, Dyreby, Jensen, Kj{\ae}rskov,
  Mikkelsen, Nielsen, and Zimmermann]{andersen2015bandwidth}
Bjarke Andersen, Jakob Dyreby, Brian Jensen, Frederik~Holmelund Kj{\ae}rskov,
  Ole~Lodahl Mikkelsen, Peter~Drustrup Nielsen, and Henrik Zimmermann.
\newblock {Bandwidth Expansion} of {Narrow Band} speech using {Linear
  Prediction}.
\newblock \emph{web source}, 26, 2015.

\bibitem[Sunil and Sinha(2012)]{sunil2012exploration}
Y~Sunil and R~Sinha.
\newblock Exploration of class specific {ABWE} for robust children's {ASR}
  under mismatched condition.
\newblock In \emph{Proceedings International Conference on Signal Processing
  and Communications (SPCOM)}, pages 1--5. IEEE, 2012.

\bibitem[Kim et~al.(2008)Kim, Lee, and Kang]{kim2008speech}
Kyung-Tae Kim, Min-Ki Lee, and Hong-Goo Kang.
\newblock Speech bandwidth extension using temporal envelope modeling.
\newblock \emph{IEEE Signal Processing Letters}, 15:\penalty0 429--432, 2008.

\bibitem[Sadasivan et~al.(2016)Sadasivan, Mukherjee, and
  Seelamantula]{sadasivan2016joint}
Jishnu Sadasivan, Subhadip Mukherjee, and Chandra~Sekhar Seelamantula.
\newblock Joint dictionary training for bandwidth extension of speech signals.
\newblock In \emph{Proceedings International Conference on Acoustics, Speech
  and Signal Processing (ICASSP)}, pages 5925--5929. IEEE, 2016.

\bibitem[Bin et~al.(2015)Bin, Jianhua, Zhengqi, Ya, Bukhari,
  et~al.]{bin2015novel}
Liu Bin, Tao Jianhua, Wen Zhengqi, Li~Ya, Danish Bukhari, et~al.
\newblock A novel method of artificial bandwidth extension using deep
  architecture.
\newblock 2015.

\bibitem[Bachhav et~al.(2017)Bachhav, Todisco, Mossi, Beaugeant, and
  Evans]{bachhav2017artificial}
Pramod~B Bachhav, Massimiliano Todisco, Moctar Mossi, Christophe Beaugeant, and
  Nicholas Evans.
\newblock Artificial bandwidth extension using the constant-{Q} transform.
\newblock In \emph{Proceedings International Conference on Acoustics, Speech
  and Signal Processing (ICASSP)}, pages 5550--5554. IEEE, 2017.

\bibitem[Sunil and Sinha(2014)]{sunil2014sparse}
Y~Sunil and Rohit Sinha.
\newblock Sparse representation based approach to artificial bandwidth
  extension of speech.
\newblock In \emph{2014 International Conference on Signal Processing and
  Communications (SPCOM)}, pages 1--5. IEEE, 2014.

\bibitem[Unno and McCree(2005)]{unno2005robust}
Takahiro Unno and Alan McCree.
\newblock A robust narrowband to wideband extension system featuring enhanced
  codebook mapping.
\newblock In \emph{Proceedings IEEE International Conference on Acoustics,
  Speech, and Signal Processing, 2005.}, volume~1, pages I--805. IEEE, 2005.

\bibitem[Pulakka et~al.(2011)Pulakka, Remes, Palom{\"a}ki, Kurimo, and
  Alku]{pulakka2011speech}
Hannu Pulakka, Ulpu Remes, Kalle Palom{\"a}ki, Mikko Kurimo, and Paavo Alku.
\newblock Speech bandwidth extension using gaussian mixture model-based
  estimation of the highband mel spectrum.
\newblock In \emph{Proceedings IEEE International Conference on Acoustics,
  Speech and Signal Processing (ICASSP)}, pages 5100--5103. IEEE, 2011.

\bibitem[Nour-Eldin and Kabal(2011)]{nour2011memory}
Amr~H Nour-Eldin and Peter Kabal.
\newblock Memory-based approximation of the gaussian mixture model framework
  for bandwidth extension of narrowband speech.
\newblock In \emph{Proceedings Twelfth Annual Conference of the International
  Speech Communication Association}, 2011.

\bibitem[Ohtani et~al.(2014)Ohtani, Tamura, Morita, and Akamine]{ohtani2014gmm}
Yamato Ohtani, Masatsune Tamura, Masahiro Morita, and Masami Akamine.
\newblock Gmm-based bandwidth extension using sub-band basis spectrum model.
\newblock In \emph{Proceedings Fifteenth Annual Conference of the International
  Speech Communication Association}, 2014.

\bibitem[Li et~al.(2018)Li, Villette, Ramadas, and Sinder]{li2018speech}
Sen Li, St{\'e}phane Villette, Pravin Ramadas, and Daniel~J Sinder.
\newblock Speech bandwidth extension using generative adversarial networks.
\newblock In \emph{Proceedings IEEE International Conference on Acoustics,
  Speech and Signal Processing (ICASSP)}, pages 5029--5033. IEEE, 2018.

\bibitem[Seltzer and Acero(2007)]{seltzer2007training}
Michael~L Seltzer and Alex Acero.
\newblock Training wideband acoustic models using mixed-bandwidth training data
  for speech recognition.
\newblock \emph{IEEE Transactions on Audio, Speech, and Language Processing},
  15\penalty0 (1):\penalty0 235--245, 2007.

\bibitem[Song and Martynovich(2009)]{song2009study}
Geun-Bae Song and Pavel Martynovich.
\newblock A study of hmm-based bandwidth extension of speech signals.
\newblock \emph{Signal Processing}, 89\penalty0 (10):\penalty0 2036--2044,
  2009.

\bibitem[Ya{\u{g}}L{\i} et~al.(2013)Ya{\u{g}}L{\i}, Turan, and
  Erzin]{yaugli2013artificial}
Can Ya{\u{g}}L{\i}, MA~Tu{\u{g}}Tekin Turan, and Engin Erzin.
\newblock Artificial bandwidth extension of spectral envelope along a viterbi
  path.
\newblock \emph{Speech Communication}, 55\penalty0 (1):\penalty0 111--118,
  2013.

\bibitem[Hinton et~al.(2012)Hinton, Deng, Yu, Dahl, Mohamed, Jaitly, Senior,
  Vanhoucke, Nguyen, Kingsbury, et~al.]{hinton2012deep}
Geoffrey Hinton, Li~Deng, Dong Yu, George Dahl, Abdel-rahman Mohamed, Navdeep
  Jaitly, Andrew Senior, Vincent Vanhoucke, Patrick Nguyen, Brian Kingsbury,
  et~al.
\newblock Deep neural networks for acoustic modeling in speech recognition.
\newblock \emph{IEEE Signal processing magazine}, 29, 2012.

\bibitem[Xu et~al.(2014)Xu, Du, Dai, and Lee]{xu2014experimental}
Yong Xu, Jun Du, Li-Rong Dai, and Chin-Hui Lee.
\newblock An experimental study on speech enhancement based on deep neural
  networks.
\newblock \emph{IEEE Signal processing letters}, 21\penalty0 (1):\penalty0
  65--68, 2014.

\bibitem[Wang et~al.(2015)Wang, Zhao, Liu, Li, and Kuang]{wang2015speech}
Yingxue Wang, Shenghui Zhao, Wenbo Liu, Ming Li, and Jingming Kuang.
\newblock Speech bandwidth expansion based on deep neural networks.
\newblock In \emph{Proceedings Sixteenth Annual Conference of the International
  Speech Communication Association}, 2015.

\bibitem[Abel and Fingscheidt(2017)]{abel2017dnn}
Johannes Abel and Tim Fingscheidt.
\newblock A dnn regression approach to speech enhancement by artificial
  bandwidth extension.
\newblock In \emph{2017 IEEE Workshop on Applications of Signal Processing to
  Audio and Acoustics (WASPAA)}, pages 219--223. IEEE, 2017.

\bibitem[Marelli and Balazs(2010)]{marelli2010pole}
Dami{\'a}n Marelli and Peter Balazs.
\newblock On pole-zero model estimation methods minimizing a logarithmic
  criterion for speech analysis.
\newblock \emph{IEEE Transactions on Audio, Speech, and Language Processing},
  18\penalty0 (2):\penalty0 237--248, 2010.

\bibitem[Chen and Francis(1995{\natexlab{a}})]{chen1995design}
Tongwen Chen and Bruce~A Francis.
\newblock Design of multirate filter banks by {$H^{\infty}$} optimization.
\newblock \emph{IEEE Transactions on Signal Processing}, 43\penalty0
  (12):\penalty0 2822--2830, 1995{\natexlab{a}}.

\bibitem[Yamamoto et~al.(2012)Yamamoto, Nagahara, and
  Khargonekar]{yamamoto2012signal}
Yutaka Yamamoto, Masaaki Nagahara, and Pramod~P Khargonekar.
\newblock {Signal Reconstruction via {$H^{\infty}$} Sampled-Data Control Theory
  Beyond the Shannon Paradigm}.
\newblock \emph{IEEE Transactions on Signal Processing}, 60\penalty0
  (2):\penalty0 613--625, 2012.

\bibitem[Shaked and Theodor(1992)]{shaked1992h}
Uri Shaked and Yahali Theodor.
\newblock {$H^\infty$} optimal estimation: a tutorial.
\newblock In \emph{Proceedings 31st IEEE Conference on Decision and Control},
  pages 2278--2286. IEEE, 1992.

\bibitem[Gupta and Shekhawat(2019{\natexlab{a}})]{gupta2019artificial2}
Deepika Gupta and Hanumant~Singh Shekhawat.
\newblock Artificial bandwidth extension using {$H^\infty$} optimization.
\newblock \emph{Proc. Interspeech 2019}, pages 3421--3425, 2019{\natexlab{a}}.

\bibitem[Nour-Eldin and Kabal(2008)]{nour2008mel}
Amr~H Nour-Eldin and Peter Kabal.
\newblock Mel-frequency cepstral coefficient-based bandwidth extension of
  narrowband speech.
\newblock In \emph{Proceedings Ninth Annual Conference of the International
  Speech Communication Association}, 2008.

\bibitem[Abel et~al.(2018)Abel, Strake, and Fingscheidt]{abel2018simple}
Johannes Abel, Maximilian Strake, and Tim Fingscheidt.
\newblock A simple cepstral domain dnn approach to artificial speech bandwidth
  extension.
\newblock In \emph{Proceedings IEEE International Conference on Acoustics,
  Speech and Signal Processing (ICASSP)}, pages 5469--5473. IEEE, 2018.

\bibitem[Loizou(2007)]{loizou2007speech}
Philipos~C Loizou.
\newblock \emph{Speech enhancement: theory and practice}.
\newblock CRC press, 2nd edition, 2007.
\newblock ISBN 13: 978-1-4665-0422-6.

\bibitem[Makhoul(1975)]{makhoul1975linear}
John Makhoul.
\newblock Linear prediction: A tutorial review.
\newblock \emph{in Proceedings IEEE}, 63\penalty0 (4):\penalty0 561--580, 1975.

\bibitem[Markel and Gray(2013)]{markel2013linear}
John~D Markel and AH~Jr Gray.
\newblock \emph{Linear prediction of speech}, volume~12.
\newblock Springer Science \& Business Media, 2013.

\bibitem[Chen and Francis(1995{\natexlab{b}})]{chen2012optimal}
Tongwen Chen and Bruce~A Francis.
\newblock \emph{Optimal sampled-data control systems}, volume 124.
\newblock Springer, 1995{\natexlab{b}}.

\bibitem[Gupta and Shekhawat(2019{\natexlab{b}})]{gupta2019artificial}
Deepika Gupta and HS~Shekhawat.
\newblock Artificial bandwidth extension using {$H^\infty$} optimization and
  speech production model.
\newblock In \emph{2019 29th International Conference Radioelektronika
  (RADIOELEKTRONIKA)}, pages 1--6. IEEE, 2019{\natexlab{b}}.

\bibitem[Aida-Zade et~al.(2006)Aida-Zade, Ardil, and
  Rustamov]{aida2006investigation}
KR~Aida-Zade, C~Ardil, and SS~Rustamov.
\newblock Investigation of combined use of mfcc and lpc features in speech
  recognition systems.
\newblock \emph{World Academy of Science, Engineering and Technology},
  19:\penalty0 74--80, 2006.

\bibitem[Garofolo(1993)]{garofolo1993timit}
John~S Garofolo.
\newblock Timit acoustic phonetic continuous speech corpus.
\newblock \emph{Linguistic Data Consortium}, 1993.

\bibitem[Larcher et~al.(2014)Larcher, Lee, Ma, and Li]{larcher2014text}
Anthony Larcher, Kong~Aik Lee, Bin Ma, and Haizhou Li.
\newblock Text-dependent speaker verification: Classifiers, databases and
  rsr2015.
\newblock \emph{Speech Communication}, 60:\penalty0 56--77, 2014.

\bibitem[Hu and Loizou(2008)]{hu2008evaluation}
Yi~Hu and Philipos~C Loizou.
\newblock Evaluation of objective quality measures for speech enhancement.
\newblock \emph{IEEE Transactions on Audio, Speech, and Language processing},
  16\penalty0 (1):\penalty0 229--238, 2008.

\bibitem[itu(2003)]{itu2003862}
{ITU-T, Recommendation P862.1: Mapping function for transforming P. 862 raw
  result scores to MOS-LQO}.
\newblock \emph{International Telecommunication Union, Geneva, Switzerland},
  2003.

\bibitem[Malfait et~al.(2006)Malfait, Berger, and Kastner]{malfait2006p}
Ludovic Malfait, Jens Berger, and Martin Kastner.
\newblock {P. 563—The ITU-T standard for single-ended speech quality
  assessment}.
\newblock \emph{IEEE Transactions on Audio, Speech, and Language Processing},
  14\penalty0 (6):\penalty0 1924--1934, 2006.

\bibitem[Kingma and Ba(2014)]{kingma2014adam}
Diederik~P Kingma and Jimmy Ba.
\newblock Adam: A method for stochastic optimization.
\newblock \emph{arXiv preprint arXiv:1412.6980}, 2014.

\bibitem[Kain and Macon(1998)]{kain1998spectral}
Alexander Kain and Michael~W Macon.
\newblock Spectral voice conversion for text-to-speech synthesis.
\newblock In \emph{Proceedings IEEE International Conference on Acoustics,
  Speech, and Signal Processing}, volume~1, pages 285--288. IEEE, 1998.

\bibitem[Adiga and Prasanna(2015)]{adiga2015detection}
Nagaraj Adiga and SRM Prasanna.
\newblock Detection of glottal activity using different attributes of source
  information.
\newblock \emph{IEEE Signal Processing Letters}, 22\penalty0 (11):\penalty0
  2107--2111, 2015.

\bibitem[rec(1996)]{rec1996p}
{ITU-T, Recommendation P.800: Methods for subjective determination of
  transmission quality}.
\newblock \emph{International Telecommunication Union, Geneva}, page~22, 1996.

\bibitem[Chiang and Safonov(1997)]{chiang1997matlab}
Richard~Y Chiang and Michael~G Safonov.
\newblock \emph{{MATLAB : Robust Control Toolbox User's Guide}}.
\newblock Math Works, 1997.

\end{thebibliography}






\end{document}